\documentclass{article}

\PassOptionsToPackage{numbers, compress}{natbib}

\usepackage[preprint]{neurips_2023}

\usepackage[utf8]{inputenc} 
\usepackage[T1]{fontenc}    
\usepackage{hyperref}       
\usepackage{url}            
\usepackage{booktabs}       
\usepackage{amsfonts}       
\usepackage{amsmath}        
\usepackage{bbm}
\usepackage{nicefrac}       
\usepackage{microtype}      
\usepackage{xcolor}         
\usepackage{adjustbox}

\graphicspath{ {images/} }

\title{Systematic Analysis of Music Representations \protect\\ from BERT}

\author{
  Sangjun Han \\
  LG AI Research \\
  \texttt{sj.han@lgresearch.ai} \\
  \And
  Hyeongrae Ihm \\
  LG AI Research \\
  \texttt{hrim@lgresearch.ai} \\
  \And
  Woohyung Lim \\
  LG AI Research \\
  \texttt{w.lim@lgresearch.ai} \\
}

\begin{document}

\maketitle

\begin{abstract}
There have been numerous attempts to represent raw data as numerical vectors that effectively capture semantic and contextual information.
However, in the field of symbolic music, previous works have attempted to validate their music embeddings by observing the performance improvement of various fine-tuning tasks.
In this work, we directly analyze embeddings from BERT and BERT with contrastive learning trained on bar-level MIDI, inspecting their musical information that can be obtained from MIDI events.
We observe that the embeddings exhibit distinct characteristics of information depending on the contrastive objectives and the choice of layers.
Our code is available at https://github.com/sjhan91/MusicBERT.

\end{abstract}

\section{Introduction}

Music consists of many repetitive components from motifs to phrases, and they have been conceptualized as forms of musical knowledge or atmosphere that humans are capable of understanding.
For instance, at the note level, performing successive multiple notes can convey harmonies and rhythmic dynamics for a short time.
At the bar-level, the performance can be expressed in chords, with chords being arranged in relationships among various bars.
At the song level, several features can serve as an overview of the composition, including played instruments, tempo, and genre.
Our focus is to understand bar-level symbolic music since it provides versatile capabilities for music analysis such as estimating musical similarity, extracting chords, and comprehending the whole structure of music.

Triggered by the field of natural language processing, there have been numerous attempts to represent raw data as numerical vectors that effectively capture semantic and contextual information.
Based on the Transformer blocks \citep{vaswani2017attention}, text embeddings can be extracted from encoder-only designs that incorporate bidirectional context \citep{devlin-etal-2019-bert}, decoder-only designs that facilitate text sequence generation \citep{radford2018improving}, and encoder-decoder designs that combine both functionalities \citep{ni2021sentence}.
In the speech, wav2vec series (wav2vec 2.0 \citep{baevski2020wav2vec}, HuBERT \citep{hsu2021hubert}, and vq-wav2vec \citep{Baevski2020vq-wav2vec}) have applied BERT \citep{devlin-etal-2019-bert} for speech representations with embedding discretization.
Also in computer vision, Vision Transformer (ViT) \citep{dosovitskiy2021an} has transformed input images into multiple grid patches and introduced the use of class token embeddings for the classification problem.

After suggested in \citep{oore2020time}, event-based representation of MIDI for machine learning has become widespread.
In this approach, each event token serves as an indicator of a specific musical action such as event changes of pitch, time shift, or velocity.
As in the following other research domains, the process of tokenization for MIDI enables us to utilize Transformer models to capture contextual information effectively.
MIDIBERT-Piano \citep{chou2021midibert} has adopted a super token-level masking strategy for BERT pre-training and demonstrated promising performance in fine-tuning performance across four tasks.
Similarly, MusicBERT \citep{zeng-etal-2021-musicbert} has proposed an efficient concept of super-token, employing a bar-level masking strategy for BERT. 
MuseBERT \citep{wang2021musebert} also has adopted BERT model, but factorized MIDI representations into attribute sets and event relation matrix.
The aforementioned previous works have attempted to validate their music embeddings by observing the performance improvement of various fine-tuning tasks.
However, when dealing with music tasks at the bar-level (\emph{e.g.} music information retrieval), it is more desirable to evaluate the embeddings based on their association with musical properties and semantics.

In this work, we directly analyze BERT embeddings trained on bar-level MIDI by inspecting their musical information that can be obtained from MIDI events.
Additionally, we compare BERT models that employ different contrastive objectives.
These models are trained with BERT loss and contrastive loss at the same time, enabling them to take into account longer contexts among bars.
This ensures that the models can be effectively adjusted to the user's intention or downstream tasks \citep{gao-etal-2021-simcse, kim-etal-2021-self, su2021improving}.
Our results demonstrate that BERT embeddings can capture important musical features and semantic information in the bar-level MIDI.
Furthermore, we observe that the embeddings exhibit distinct characteristics of information depending on the contrastive objectives and the choice of layers.

\section{Method}

\begin{figure}
  \centering
  \includegraphics[width=10cm]{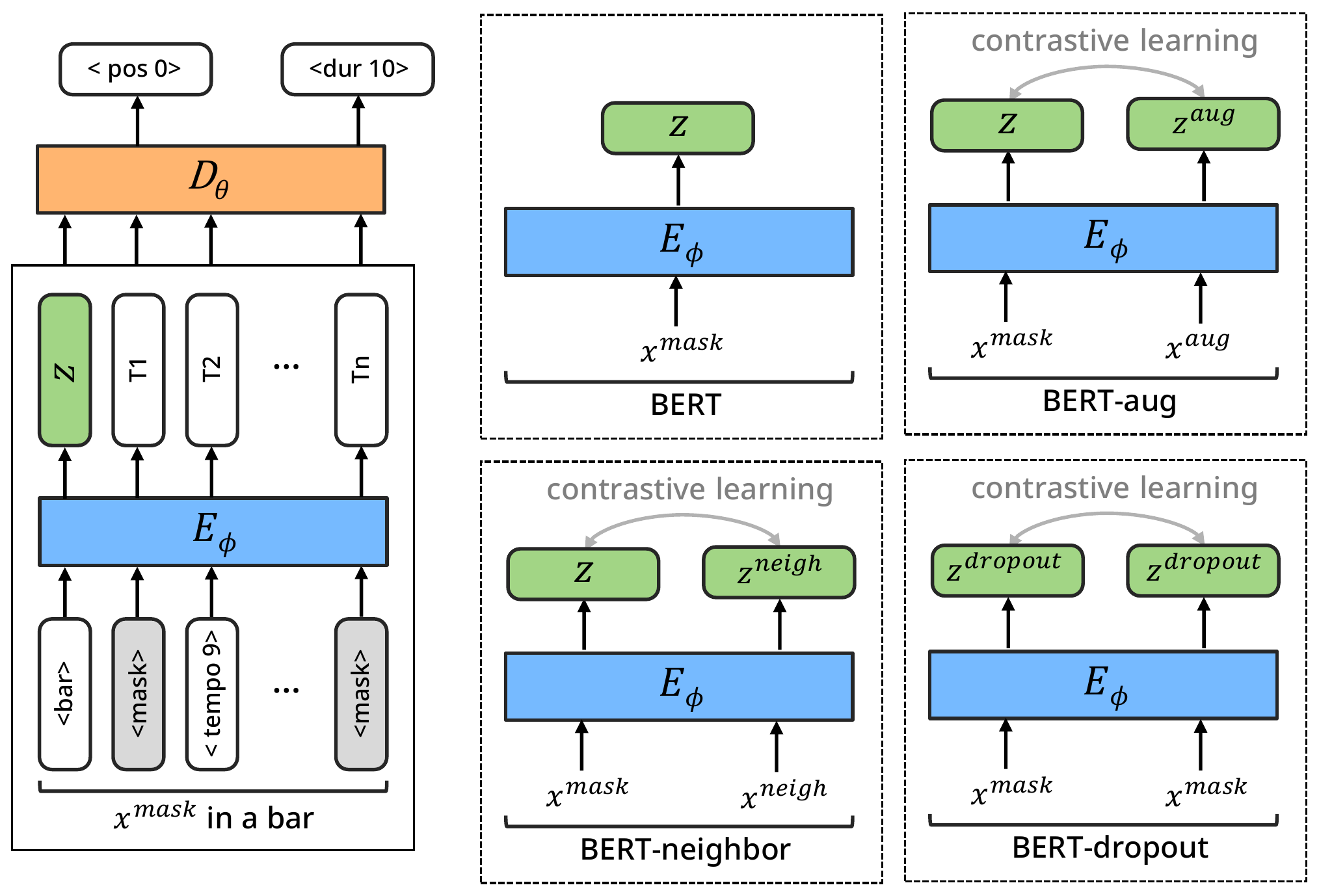} \\
  \caption{Left: The basic structure of BERT with masked prediction, Right: Various contrastive objectives that are trained in conjunction with masked prediction.}
\end{figure}

In this section, we introduce the process of data preparation, several design concepts for music embedding models, and the evaluation protocol.

\paragraph{Data Preparation}
Among symbolic music datasets, Lakh MIDI Dataset (LMD) \citep{raffel2016learning} is widely used since it comprises 176,581 MIDI files spanning diverse genres and tracks.
We convert each MIDI from LMD into REMI+ representation \citep{von2022figaro}, an extended version of REMI \citep{huang2020pop} that enables the expression of multiple tracks.
Our vocabulary contains 556 events for 8 categories; 1 <bar>, 32 <tempo>, 129 <instrument>, 128 <pitch>, 128 <pitch drum>, 48 <position>, 58 <duration>, and 32 <velocity>.
We adopt the same configuration for REMI+ as described in \citep{von2022figaro}.
In the end, we collect a total of 9,971,616 bars from the LMD.

\paragraph{Model Descriptions}
Our embedding models are following \textbf{BERT\textsubscript{base}} model configuration (the number of layers=12, the hidden size=768, the number of self-attention heads=12).
During the training process, we utilize masked language modeling loss (MLM loss) which involves masking a portion of input tokens and predicting those tokens.
At each iteration, 15\% of input tokens are selected, among which 80\% of the tokens are masked, 10\% of the tokens are randomly replaced, and the remaining 10\% of tokens remain unchanged.
We remove the next sentence prediction task from the original BERT.

We introduce three variant models derived from BERT; BERT-aug, BERT-neighbor, and BERT-dropout.
These models are trained MLM loss and contrastive loss (NT-Xent loss) formulated at SimCLR \citep{chen2020simple} simultaneously.
For a minibatch $N$, the NT-Xent loss can be defined as

\begin{equation}
L_{NT-Xent}(z^{\prime}, z^{\prime\prime}) = -log \frac{exp(sim(z^{\prime},z^{\prime\prime}) / \tau)} {\sum_{k=1}^{2N} \mathbbm{1}_{[z_{k} \neq z^{\prime}]} exp(sim(z^{\prime},z_{k}) / \tau)}
\end{equation}

where $z^{\prime}$ and $z^{\prime\prime}$ represent a positive pair that is semantically identical, while $z_{k}$ is sampled from a negative set.
Given a $N$ batch, we generate a $N$ new batch of positive views using predefined functions (\emph{e.g.} augmentation), resulting in 2$N$ samples in a batch.
When considering a single positive pair, the remaining $2(N-1)$ samples are regarded as the negative set.
In this context, $Sim(\cdot,\cdot)$ denotes the cosine similarity, $\mathbbm{1}$ does the indicator function, and $\tau$ does the temperature parameter.
Then, our total loss can be described as

\begin{equation}
L= L_{MLM} + \alpha \cdot L_{NT-Xent}
\end{equation}

where $\alpha$ controls the degree of NT-Xent loss.
We set $\tau$ and $\alpha$ to 0.1 for all experiments.
Unlike previous studies \citep{gao-etal-2021-simcse, kim-etal-2021-self, su2021improving}, since we train the two objectives (MLM loss and NT-Xent loss) concurrently, masked inputs ($x^{mask}$) inevitably are involved in the training process of contrastive learning as shown in Figure 1.
In other words, $x^{mask}$ can be regarded as one of the augmented views from the predefined functions.
It is analogous to Mask Contrast \citep{zhao2021self} for image representations in that both the masked view and augmented view are participating in the contrastive loss.
Below, we provide a comprehensive explanation of the design principles behind our variant models.

\begin{itemize}
  \item \textbf{BERT-aug}: To generate the positive view of samples, we apply data augmentation to original sample $x$ by shifting randomly all pitches \{-6, -5, ... , 5, 6\} and velocities \{-3, -2, ... , 2, 3\} in a sequence, resulting in $x^{aug}$.
  It still maintains the melodic contour, not undermining musical semantics.
  Similar strategies for data augmentation can be found in \citep{huang2018music, shih2022theme}.
  
  \item \textbf{BERT-neighbor}: This is motivated by NNCLR \citep{dwibedi2021little} which is a contrastive model regarding nearest neighbors of the augmented view as positives.
  In our setting, a sample $x^{neigh}$ is considered to be a neighbor of $x$ if they belong to the same music.
  
  \item \textbf{BERT-dropout}: This is motivated by SimCSE \citep{gao-etal-2021-simcse} which adopts Dropout \citep{srivastava2014dropout} as a stochastic augmentation.
  The model receives the same $x^{mask}$ twice in the forward pass and generates two different embeddings in a positive relation.
  We place the Dropout mask on attention maps and feed-forward networks in Transformer blocks and set the masking rate to 0.1.
\end{itemize}

Table 1 compares the MLM accuracy and NT-Xent accuracy of each model for the training and validation set.
Exceptionally, BERT-neighbor exhibits significant disparities between the training and validation accuracy even after adjusting the values of $\tau$ and $\alpha$.
We can speculate that neighbors within the same music do not extensively share musical information.

\begin{table}[!t]
  \caption{The training and validation accuracy of four models, indicating the masked language modeling loss (MLM) and the normalized temperature-scaled cross-entropy loss (NT-Xent).}
  \centering
  \begin{tabular}{ccc}
    \toprule
    Model  & MLM Acc  & NT-Xent Acc \\
    \midrule
    BERT  & 0.889 / 0.869  & - \\
    BERT-aug  & 0.892 / 0.869  & 1.000 / 0.918 \\
    BERT-neighbor  & 0.900 / 0.868  & 1.000 / 0.059 \\
    BERT-dropout  & 0.890 / 0.868  & 1.000 / 1.000 \\
    \bottomrule
  \end{tabular}
\end{table}

\paragraph{Evaluation Methods}

We evaluate the bar-level BERT embeddings on their alignment with human interpretable domain knowledge.
Referring to \citep{von2022figaro}, the metrics can be listed as follows; chords, groove patterns, instruments, tempo, mean velocity, mean duration, and song clustering.
The evaluation entails assessing the performance of linear probing tasks, including multi-class classification with a Ridge classifier, multi-label classification with a Ridge classifier, regression with a Ridge regressor, and clustering with K-means.
We provide a detailed explanation of each of these metrics and evaluation methods.

\begin{itemize}
  \item \textbf{Chords (C)}: As following \citep{von2022figaro, huang2020pop}, we extract chords using an adapted version of the Viterbi algorithm.
  They consist of 12 root notes and 7 qualities, resulting in a total of 84 possible chords.
  Since multiple chords can be placed on a bar, we evaluate the performance of multi-label classification for the chords.
  
  \item \textbf{Groove Patterns (GP)}: We label a position as 1 in a bar if any note is played and as 0 if no note is present.
  We evaluate the performance of multi-label classification for the groove patterns.

  \item \textbf{Instruments (I)}: We label an instrument as 1 if the instrument appears.
  We evaluate the performance of multi-label classification for the instruments.

  \item \textbf{Tempo (T)}: Tempos are quantized into 32 bins.
  We evaluate the performance of multi-class classification for the tempos.

  \item \textbf{Mean Velocity (MV)}: We compute the average of all velocity values within a bar.
  We evaluate the performance of regression for the mean velocity.

  \item \textbf{Mean Duration (MD)}: We compute the average of all duration values within a bar.
  We evaluate the performance of regression for the mean duration.

  \item \textbf{Song Clustering (SC)}: Using K-means, we compute the average of entropy based on the number of classes assigned to the bars in the same music.
  The low entropy means that the bars in the same music tend to be clustered to the same class.
  This metric can verify how shared semantics permeating through music can be extracted.
\end{itemize}

\section{Experiments}

We empirically demonstrate quantitative evaluations of musical information from the BERT-variants models.
First, we inspect the information from the last layer of Transformer blocks and the changes in the amount of information across different layers.

\paragraph{The inspection of BERT embeddings from the last layer}

\begin{table}[!t]
  \caption{The performance of linear probing using embeddings from the last layer of BERT models. Best values are marked in bold font.}
  \centering
  \resizebox{\textwidth}{!}
    {
        \begin{tabular}{cccccccc}
            \toprule
            Model  & C (\(\uparrow\))  & GP (\(\uparrow\))  & I (\(\uparrow\))  & T (\(\uparrow\))  & MV (\(\downarrow\))  & MD (\(\downarrow\))  & SC (\(\downarrow\)) \\
            \midrule
            BERT  & \textbf{0.861}  & \textbf{0.808}  & 0.764  & 0.083  & 9.169  & 1.904  & 0.252 \\
                  & \textbf{(1.933e-4)}  & \textbf{(4.430e-4)}  & (1.674e-4)  & (1.593e-4)  & (1.482e-2)  & (1.728e-2)  & (4.154e-4) \\
            BERT-aug  & 0.532  & 0.805  & \textbf{0.855}  & 0.027  & 13.093  & \textbf{1.804}  & 0.217 \\
                      & (4.058e-4)  & (4.098e-4)  & \textbf{(2.186e-4)}  & (3.661e-4)  & (1.432e-2)  & \textbf{(1.401e-2)}  & (1.442e-3) \\
            BERT-neighbor  & 0.671  & 0.768  & 0.823  & \textbf{0.946}  & \textbf{7.107}  & 2.371  & \textbf{0.072} \\
                           & (5.486e-4)  & (4.142e-4)  & (1.951e-4)  & \textbf{(2.277e-4)}  & \textbf{(1.393e-2)}  & (1.790e-2)  & \textbf{(4.454e-4)}  \\
            BERT-dropout  & 0.797  & 0.776  & 0.750  & 0.042  & 8.943  & 1.937  & 0.226 \\
                          & (3.031e-4)  & (4.789e-4)  & (2.566e-4)  & (2.967e-4)  & (1.052e-2)  & (1.702e-2)  & (8.667e-4) \\
            \bottomrule
        \end{tabular}
    }
\end{table}

\begin{table}[!b]
  \caption{The best performance of linear probing using embeddings across all BERT layers. We indicate the performance as (the layer number, the performance of the layer). Best values are marked in bold font.}
  \centering
  \resizebox{\textwidth}{!}
    {
        \begin{tabular}{cccccccc}
            \toprule
            Model  & C (\(\uparrow\))  & GP (\(\uparrow\))  & I (\(\uparrow\))  & T (\(\uparrow\))  & MV (\(\downarrow\))  & MD (\(\downarrow\))  & SC (\(\downarrow\)) \\
            \midrule
            BERT  & \textbf{(12, 0.861)}  & \textbf{(5, 0.863)}  & (4, 0.882)  & (1, 0.933)  & \textbf{(2, 4.905)}  & (1, 1.571)  & (3, 0.169) \\
            BERT-aug  & (1, 0.566)  & (4, 0.839)  & (12, 0.855)  & (2, 0.337)  & (2, 9.823)  & \textbf{(6, 1.453)}  & (12, 0.219) \\
            BERT-neighbor  & (8, 0.716)  & (2, 0.852)  & \textbf{(4, 0.941)}  & \textbf{(4, 0.990)}  & (2, 5.159)  & (2, 1.584)  & \textbf{(6, 0.062)} \\
            BERT-dropout  & (12, 0.797)  & (7, 0.804)  & (5, 0.807)  & (2, 0.664)  & (5, 5.985)  & (5, 1.664)  & (12, 0.226)\\
            \bottomrule
        \end{tabular}
    }
\end{table}

Table 2 reports linear probing tasks for the embeddings of the last layer in terms of four models and seven metrics.
Remarkably, the original BERT exhibits the highest performance for the chord classification.
It can be inferred that during the augmentation process (or neighbor sampling), the other models augment pitch events in positive pairs so that they obtain invariant features for the chords.
Similarly, BERT-aug shows the lowest performance for the mean velocity since it specifically modifies the velocity values during the augmentation.
BERT-dropout controls all factors so that it decreases all performance metrics compared to the best performances.

The advantage of contrastive learning lies in its ability to extract shared information in a positive view.
For BERT-aug, the mean duration is not a variable factor between bars in a positive relationship, which may improve its performance.
For BERT-neighbor, it tends to successfully classify bars into their respective songs in terms of song clustering.

The results of groove pattern, instrument, and tempo classification are inconsistent with subsequent results that analyze the performance across all BERT layers.
We supplement the explanation in the following section.

\paragraph{The inspection of BERT embeddings across all layers}
Table 3 reports the best performance and its layer number for the linear probing tasks.
Except for BERT-neighbor, the models with significantly lower association with the tempo show the performance improvement depending on the choice of layers.
The information on groove patterns and instruments from all models is distributed across all layers.
Figure 2 illustrates the layer-wise performance of the BERT-variants model.

\begin{figure}
  \centering
  \includegraphics[width=\textwidth]{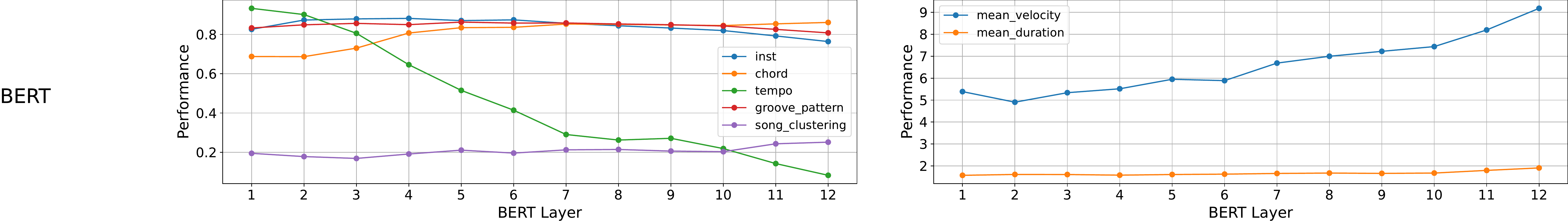} \\
  \vspace{0.25cm}
  \includegraphics[width=\textwidth]{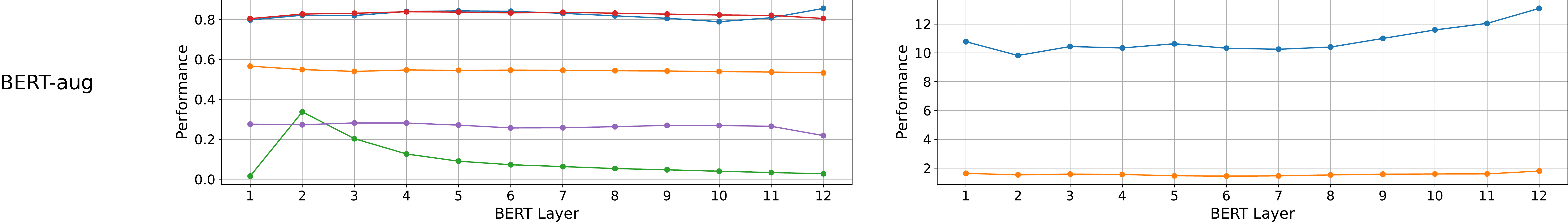} \\
  \vspace{0.25cm}
  \includegraphics[width=\textwidth]{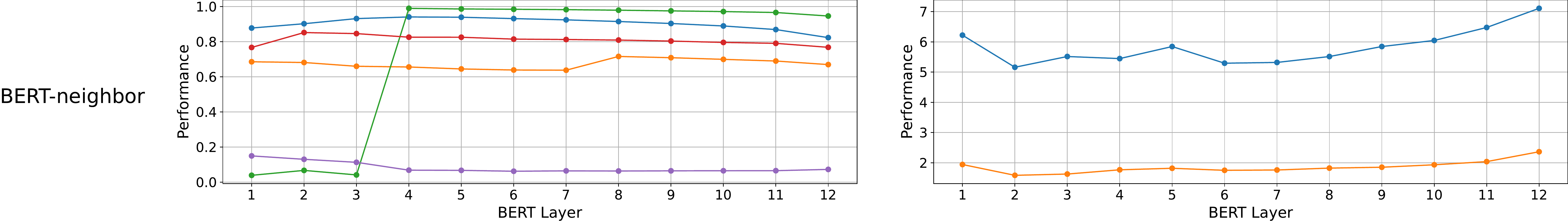} \\
  \vspace{0.25cm}
  \includegraphics[width=\textwidth]{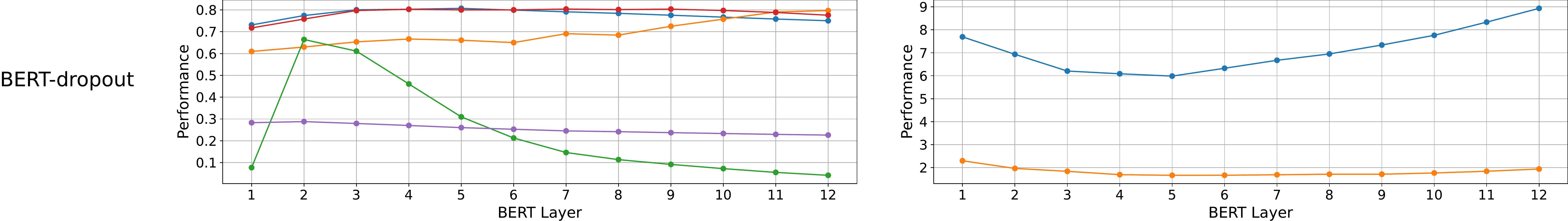}
  \caption{The layer-wise performance of the BERT, BERT-aug, BERT-neighbor, and BERT-dropout (from top to bottom).}
\end{figure}

\section{Discussion}

Several studies have analyzed BERT embeddings for various NLP tasks \citep{kim-etal-2021-self, de-vries-etal-2020-whats, jawahar-etal-2019-bert}.
They indicate that BERT does not follow the classical NLP pipeline (simple to complex) or exhibits distributed information across its layers.
Nevertheless, our research provides consistent evidence for several factors and demonstrates the effectiveness of contrastive learning.
Especially for BERT-neighbor, it can be utilized to extract a musical theme effectively.
As mentioned in \citep{de-vries-etal-2020-whats}, the integration of information from various layers will be important for enhancing the quality of information.

In this paper, we perform a systemic analysis of bar-level music embeddings from BERT and BERT with contrastive learning models.
For seven metrics, our linear probing tasks can assess the amount of musical information, revealing the effectiveness of specific models for certain metrics.
The bar-level embedding models will contribute to various musical tasks, including obtaining chord extraction, music similarity analysis, and music structure understanding.

\newpage
\bibliographystyle{unsrtnat}
\bibliography{ref}


\end{document}